\documentclass[usenatbib,twocolumn]{mn2e}
\usepackage{times, graphicx, amsfonts, amssymb}

\pubyear{2009}

\def\beq{\begin{equation}}
\def\eeq{\end{equation}} 
\def\baq{\begin{eqnarray}}
\def\eaq{\end{eqnarray}}

\def\p3m{P$^3$M}
\def\ap3m{AP$^3$M}

\def\rnl{r_\mathrm{nl}}

\def\h1{HI}
\def\omegah1{\Omega_{\h1}}
\def\ph1{P_{_{\h1}}}
\def\ph1k{P_{_{\h1}}(k)}
\def\dh1k{\Delta^2_{_{\h1}}(k)}

\def\neff{n_\mathrm{eff}}

\title[Mass function of scale invariant models]{Mass function of haloes: scale
  invariant models} 

\author[Bagla, Khandai and Kulkarni] {J.~S.\ Bagla, Nishikanta Khandai and
  Girish Kulkarni\\ 
Harish-Chandra Research Institute, Chhatnag Road, Jhunsi, \\
Allahabad 211019, INDIA\\ E-Mail: jasjeet, nishi, girish@hri.res.in}
\pubyear{2009}

\label{firstpage}

\def\LaTeX{L\kern-.36em\raise.3ex\hbox{a}\kern-.15em
    T\kern-.1667em\lower.7ex\hbox{E}\kern-.125emX}

\pagerange{\pageref{firstpage}--\pageref{lastpage}} 

\begin{document}

\maketitle

\begin{abstract}
Press-Schechter theory gives a simple, approximate functional
form of the mass function of dark matter haloes.
Sheth and Tormen (ST) refined this mass function to give an improved
analytical fit to results of $N$-body simulations.
These forms of the halo mass function are universal (independent of
cosmology and power spectrum) when scaled in suitable variables.
Using large suites of LCDM $N$-body simulations, studies in the last
few years have shown that this universality is only approximate.
We explore whether some of the deviations from universality can be attributed
to the power spectrum by computing the mass function in $N$-body simulations
of various scale-free models in an Einstein-de~Sitter cosmology. 
This choice of cosmology does not introduce any scale into the problem.
These models have the advantage of being self-similar, hence stringent checks
can be imposed while running these simulations.  
This set of numerical experiments is designed to isolate any power spectrum
dependent departures from universality of mass functions. 
We show explicitly that the best fit ST parameters have a clear dependence on
power spectrum.  
Our results also indicate that an improved analytical theory with more 
parameters is required in order to provide better fits to the 
mass function. 
\end{abstract}




\section{Introduction}

The halo mass function describes number density of dark matter haloes
of a given mass in a given cosmology, and is an essential input for a diverse
set of tools used for making theoretical predictions.
The halo model of large scale structure, for example, is based on the theory
of mass functions \citep{2002PhR...372....1C}.
Accurate knowledge of the mass function is important for several
cosmological applications, including semi-analytic theories of galaxy
formation \citep{1991ApJ...379...52W}; constraints on cosmological
parameters using galaxy cluster abundance \citep{2003ApJ...585..603M},
merger rates for haloes \citep{1994MNRAS.271..676L, 2001MNRAS.325.1053C},
gravitational lensing \citep{1998A&A...330....1B} and constraints on
non-Gaussianity in the primordial power spectrum
\citep{2005JCAP...10..010B, 2008arXiv0811.4176P} of matter perturbation.

It is possible to develop the theory of mass functions in a manner that makes
no reference to the details of the cosmological model or the power spectrum of
fluctuations. 
That is, we expect the mass function to take a universal form, when scaled
appropriately. 
Simple theoretical arguments have been used to obtain this universal
functional form of the mass function \citep{1974ApJ...187..425P,
  1991ApJ...379..440B, 2001MNRAS.323....1S}.  
\citet{1991ApJ...379..440B} and \citet{2001MNRAS.323....1S} used the
excursion set theory to derive the mass function. 
Much work has also been done to determine the extent to which this
form is consistent with results from N-body simulations
\citep{2001MNRAS.321..372J, 2002ApJS..143..241W, 2003MNRAS.346..565R,
  2006ApJ...646..881W, 2007MNRAS.374....2R, 2007ApJ...671.1160L,
  2008MNRAS.385.2025C, 2008ApJ...688..709T} 
with the conclusion that the agreement is fairly good.
Recent comparisons with very large N-Body simulations also provide hints that
the form of the mass function is not universal.

The Press-Schechter mass function \citep{1974ApJ...187..425P} is based on the 
spherical collapse model \citep{1972ApJ...176....1G} and the 
ansatz that the mass in collapsed objects is related to the volume
with density above a certain threshold.
The shape of the mass function agrees with numerical results qualitatively: at
a quantitative level there are deviations at the low mass
and the high mass ends \citep{1988MNRAS.235..715E, 2001MNRAS.321..372J}. 
Improvements to the Press-Schechter mass function have been made to
overcome this limitation.  
In particular, the Sheth-Tormen mass function is based on the more
realistic ellipsoidal collapse model \citep{1999MNRAS.308..119S,
  2001MNRAS.323....1S} and it fits numerical results better.
These mass functions relate the abundance of haloes to the
initial density field in a universal manner, independent of cosmology
and power spectrum.
Many fitting functions with three or four fitting parameters have been
proposed.
These are based on results of simulations of the LCDM model
\citep{2001MNRAS.321..372J,2003MNRAS.346..565R, 2006ApJ...646..881W}.   

In the last few years, large N-Body simulations of the LCDM model have
demonstrated that the mass function is not universal
\citep{2002ApJS..143..241W}, and epoch dependent fitting functions have been
given for this model \citep{2007MNRAS.374....2R,2008ApJ...688..709T}. 
The results of \citet{1994MNRAS.271..676L} also show a small dependence 
of the mass function on the power spectrum but given the size of simulations 
these deviations are small. 
These studies show non-universality by noting variations in the form of mass
function with redshift or cosmology in successively larger simulations
that explore a large range in mass.
Much of the numerical work in this area during the last decade has focused on
the LCDM model, it being the model favored by observations.

It is expected that this non-universality is a result of variation of
mass function parameters either with cosmology or with the power
spectrum or both.  
Cosmology dependence is introduced by the variation in the threshold density
for collapse \citep{1993MNRAS.262..717B}.
The CDM class of models have a power spectrum of density fluctuations with a
gradually varying slope or the spectral index, $n(k)$, which decreases with
decreasing scale (increasing wavenumber $k$).
As perturbations at smaller scales collapse earlier, the effective index of
the power spectrum is small at early times and increases towards late times. 
The threshold overdensity for collapse also changes as the cosmological
constant becomes more important at late times. 
Thus the variation in mass function may be due to the shape of the power
spectrum, or cosmology, or both.
This makes the published results difficult to interpret in terms of a
theoretical model.  
It is then hard to discern any trends in non-universality that will
potentially provide a physical understanding.  
Given the number of applications of the theory of mass function like
computation of merger rates, halo formation rates, etc., it is essential to
develop a clear understanding of the origin of non-universality.
The only other option is to work with fitting functions for each of these
quantities derived from N-Body simulations.

The problem of an unclear origin of deviations of universality can be
partially addressed by studying a wider variety of models in the CDM class of
models. 
This approach has been taken by, for example, \citet{2009arXiv0903.1640N} in
the context of universality in halo mergers.  
The effect of perturbations at larger scales in terms of the tidal
field, which is relevant for ellipsoidal collapse, also changes with time due
to the variation in the slope of the power spectrum with scale.
Given that the tidal field is generated by larger scales, it is not very clear
whether a prescription based on the local index of the power spectrum alone
can provide a detailed explanation for the mass function.
Since the CDM spectra lack the simplicity of the scale-free spectra, we
approach this problem differently.  
We specifically look for departures from non-universality in the mass
function for scale-free power spectra of initial fluctuations with an
Einstein-de~Sitter background to check if the non-universal description can be
attributed to a spectrum dependence. 
Our choice of cosmology does not introduce any scale in the problem and the
threshold overdensity does not vary with time in any non-trivial manner.
Thus we can isolate the non-universality of mass functions arising from the
slope of the power spectrum.
We provide spectrum dependent fits for the parameters in the Sheth-Tormen mass
function, and show that this allows us to fit simulation data much
better. 

We start with a discussion of the basic framework of mass functions in
\S\ref{massfunction}, where we also set up the notation.
Our numerical simulations are described in \S\ref{simulations}.  
We present our analysis of the data in \S\ref{results} along with the
results.  
A discussion of their implications appears in \S\ref{discussion} and
we summarize our conclusions in \S\ref{conclusions}.

\section{The Mass Function}
\label{massfunction}

The mass function is described by the following function in the
Press-Schechter formalism: 
\beq
f(\nu)=\sqrt\frac{2}{\pi}\nu\exp(-\nu^2/2) 
\label{fps}
\eeq 
where $\nu= \delta_c/ \left(\sigma(m) D_+(z)\right)$.  
Here $\delta_c$ is the threshold overdensity for a spherically symmetric
perturbation, above which it collapses and forms a virialised halo.
It has only a weak dependence on cosmology and its value is $1.69$ for
$\Omega_0 = 1$ \citep{1972ApJ...176....1G, 1980lssu.book.....P}.
Fitting functions that describe the dependence of $\delta_c$ on
cosmology are available in the literature \citep{1993MNRAS.262..717B,
  1996MNRAS.282..263E, 1996ApJ...462..563N, 1997PThPh..97...49N,
  1998ApJ...495...80B, 2000ApJ...534..565H}.  
We take the value of $1.686$ throughout our analysis as we are working with
the Einstein-de Sitter cosmology.

The {\it rms} fluctuations of the linearly evolved density field at present is
denoted by $\sigma(m)$, smoothed with a spherical top hat filter enclosing
mass $m$ and is calculated by convolving the linear density power spectrum
$P(k)$, extrapolated to the current epoch, with the filter $W(k,m)$:  
\beq
\sigma^2(m) = \int_0^\infty \frac{dk}{k}\frac{k^3P(k)}{2\pi^2}W^2(k,m).
\eeq
Lastly, $D_+(z)$ is the growth function \citep{1977MNRAS.179..351H}.
All mass is contained in haloes in this formalism, this provides the
normalization:  
\beq 
\int_{0}^\infty \frac{1}{\nu} f(\nu)d\nu=1
\eeq 
This mass function is related to the number of haloes of a given mass
per unit comoving volume by 
\beq
\frac{dn}{d\ln m}=\frac{\bar\rho}{m}\frac{d\ln\sigma^{-1}}{d\ln
  m}f(\nu).
\label{norm}
\eeq
The dependence on cosmology and power spectrum is absorbed in $\nu$, and the
theory of mass functions can be developed without reference to the detailed
dependence of $\nu$ on the power spectrum or cosmology, so we expect the form
of Equation (\ref{fps}) to be universal. 

The Sheth-Tormen mass function is a modification to the Press-Schechter model
and is based on ellipsoidal collapse instead of spherical collapse.
It has been shown to reproduce simulation results better.
\beq 
f(\nu)= A\sqrt\frac{2q}{\pi} \left[1+(q\nu^2)^{-p}\right]
\nu\exp\left(-q\nu^2/2\right). 
\label{fst}
\eeq 
Clearly, this is also a universal form. 

\begin{table}
\caption{Details of $N$-body simulations used in this work.}
\begin{center}
\begin{tabular}{||l|l|l|l|r|r|l||}
\hline
$n$ & $N_\mathrm{box}$ & $N_\mathrm{part}$ & $\rnl^i$ & $\rnl^f$ & $\rnl^\mathrm{max}$ & $z_i$\\
\hline
$-2.5$ & $512^3$ & $512^3$ & $0.2$ & $1.0$ & $0.1$ & $36.21$\\
$-2.2$ & $512^3$ & $512^3$ & $0.5$ & $2.0$ & $2.0$ & $51.23$\\
$-2.0$ & $512^3$ & $512^3$ & $1.0$ & $4.5$ & $4.2$ & $62.80$\\
\hline
$-1.8$ & $512^3$ & $512^3$ & $2.5$ & $9.0$ & $8.5$ & $78.82$\\
$-1.5$ & $400^3$ & $400^3$ & $2.5$ & $12.0$ & $10.0$ & $103.38$\\
$-1.0$ & $400^3$ & $400^3$ & $2.5$ & $10.0$ & $22.2$ & $171.52$\\
$-0.5$ & $256^3$ & $256^3$ & $2.5$ & $12.0$ & $18.2$ & $291.53$\\
$+0.0$ & $256^3$ & $256^3$ & $2.5$ & $12.0$ & $21.2$ & $470.81$\\
\hline
\label{table_nbody_runs}
\end{tabular}
\end{center}
\end{table}

\section{Numerical Simulations}
\label{simulations}

We run a suite of models with a power law power spectrum ($P(k) = Ak^n$)
of initial fluctuations, in the range $-2.5 \geq n \geq 0.0$.
We use the Einstein-de~Sitter cosmological background where the
growing mode of perturbations $D_+(t)$ is the same as the scale factor
$a(t)$.
We used the TreePM code \citep{2009RAA.9...861} for these simulations.
The TreePM \citep{2002JApA...23..185B, 2003NewA....8..665B}
is a hybrid N-Body method which improves the accuracy and performance
of the Barnes-Hut (BH) Tree method \citep{1986Natur.324..446B} by
combining it with the PM method \citep{1983ApJ...270..390M,
  1983MNRAS.204..891K, 1985A&A...144..413B, 1985ApJ...299....1B,
  1988csup.book.....H, 1997Prama..49..161B, 2005NewA...10..393M}.
The TreePM method explicitly breaks the potential into a short-range
and a long-range component at a scale $r_s$: the PM method is used to
calculate long-range force and the short-range force is computed using
the BH Tree method.
Use of the BH Tree for short-range force calculation enhances the force
resolution as compared to the PM method.

The mean interparticle separation between particles in the simulations
used here is $l_\mathrm{mean} = 1.0 $ in units of the grid-size used
for the PM part of the force calculation.
In our notation this is also cube root of the ratio of simulation volume
$N_\mathrm{box}^3$ to the total number of particles $N_\mathrm{part}$.

\begin{figure*}
  \begin{center}
    \begin{tabular}{cc}
      \includegraphics[width=2.85truein]{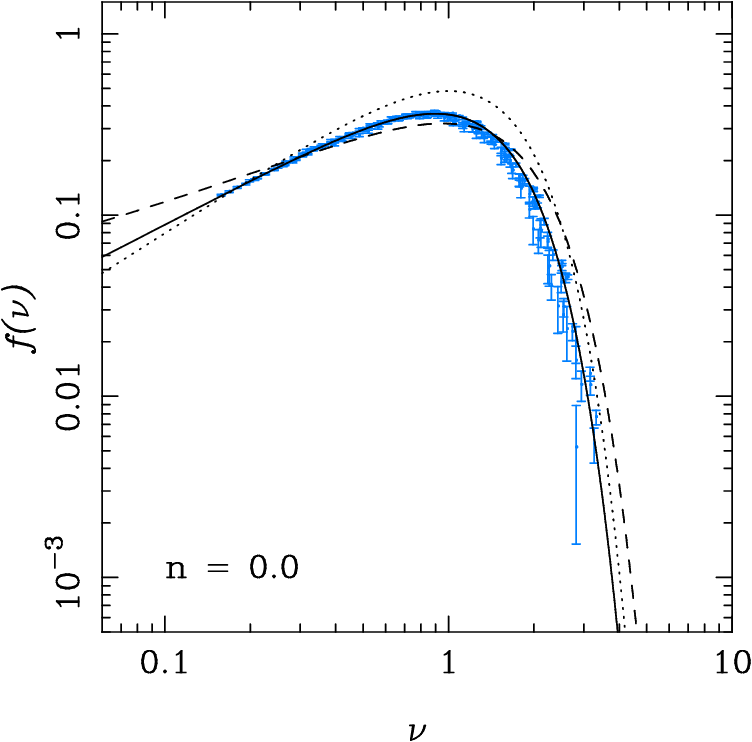} &
      \includegraphics[width=2.85truein]{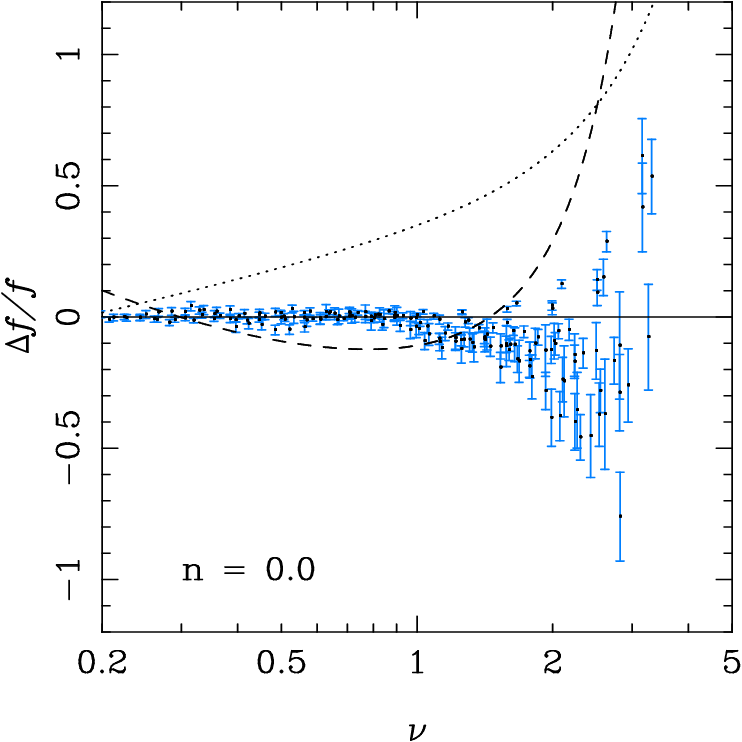} \\
      \includegraphics[width=2.85truein]{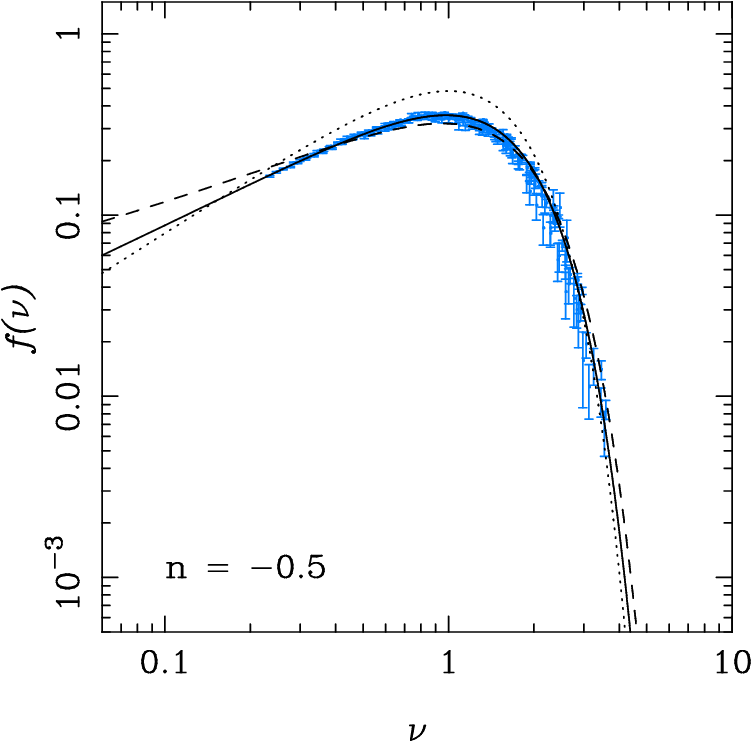} &
      \includegraphics[width=2.85truein]{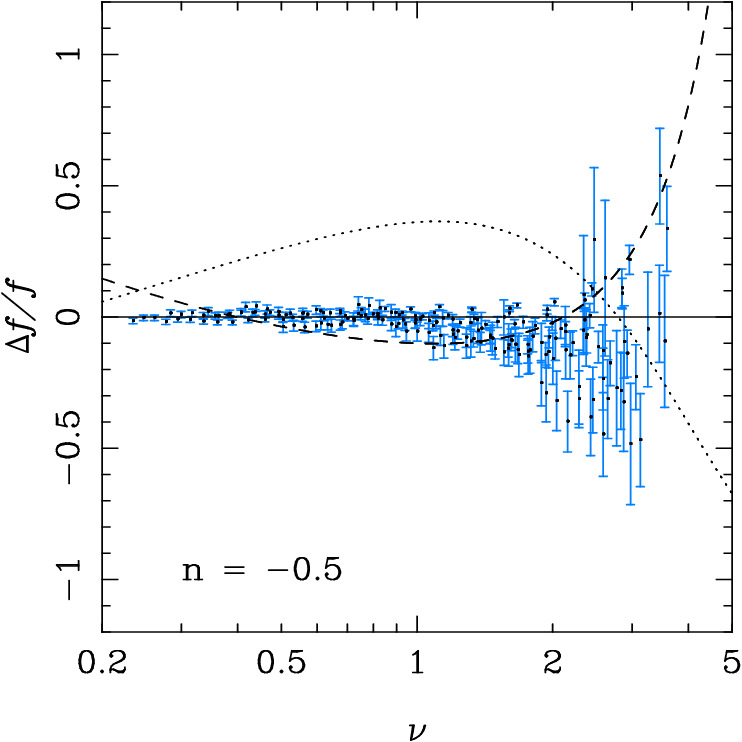} \\
      \includegraphics[width=2.85truein]{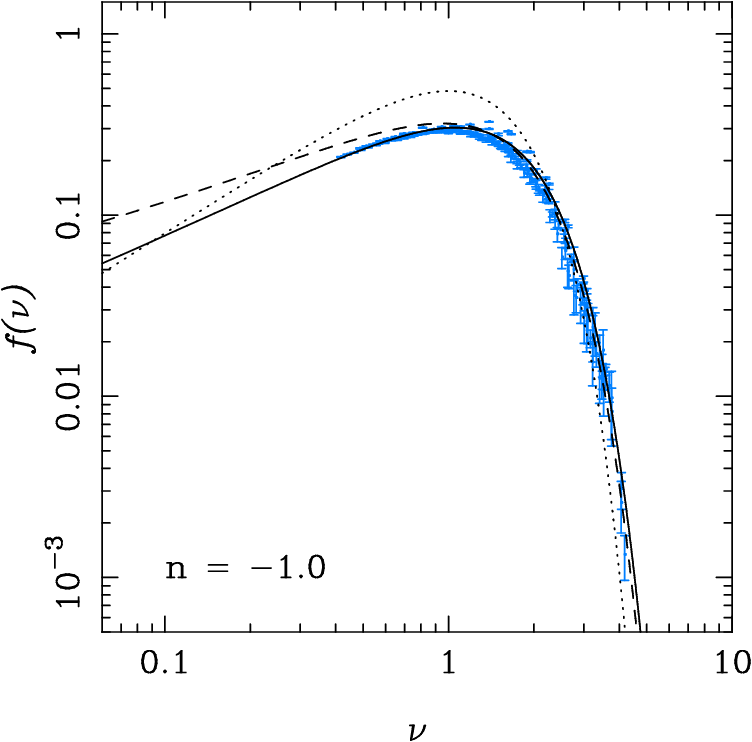} &
      \includegraphics[width=2.85truein]{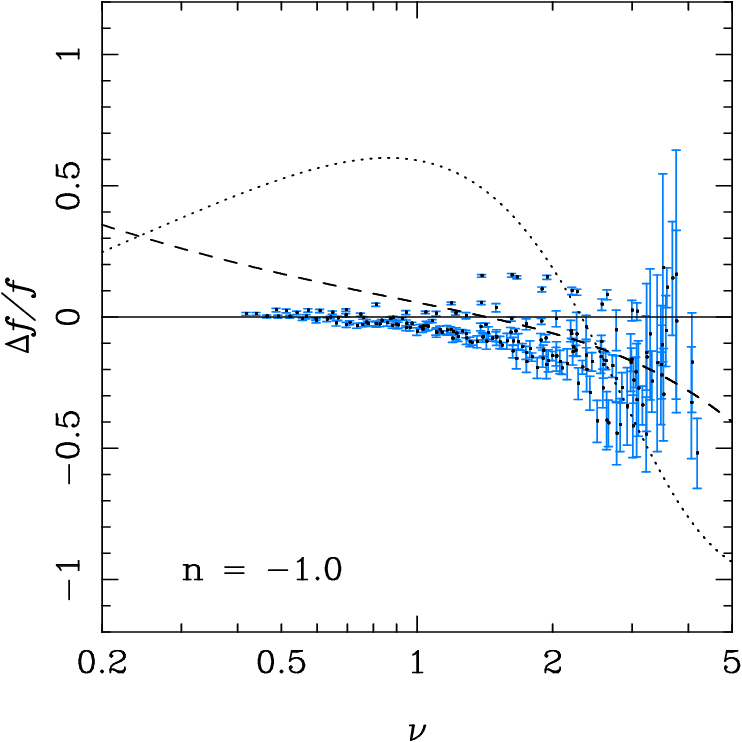} \\
    \end{tabular}
  \end{center}
  \caption{Mass functions from our simulations (left column) and their
    residuals after fitting a Sheth-Tormen form (right column).  Rows
    1, 2 and 3 are for indices $n=0.0$, $-0.5$, $-1.0$.  Solid black
    line is our best fit Sheth-Tormen curve.  Dashed line and dotted
    line show the standard Sheth-Tormen and Press-Schechter curves
    respectively.}
  \label{fig_best_fit1}
\end{figure*}
\begin{figure*}
  \begin{center}
    \begin{tabular}{cc}
      \includegraphics[width=2.85truein]{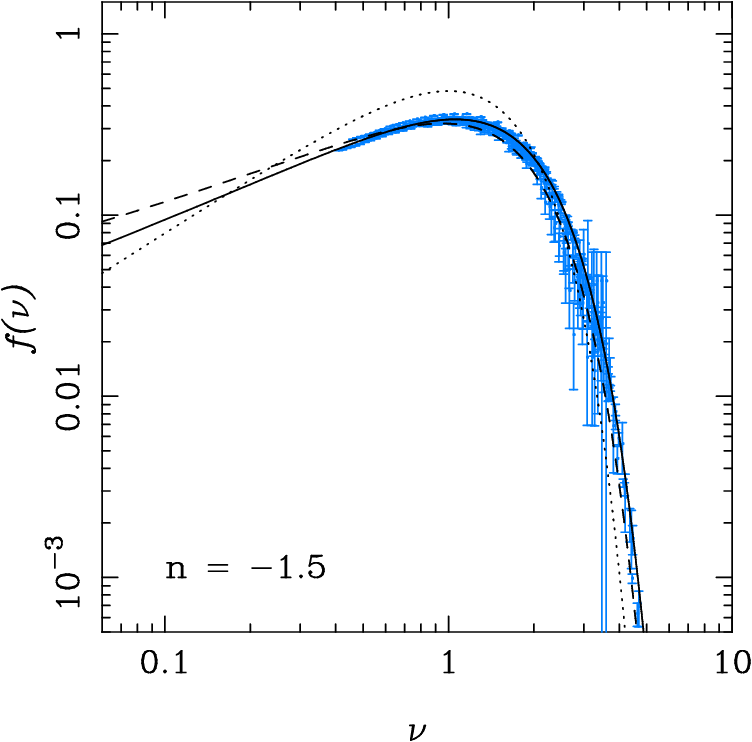} &
      \includegraphics[width=2.85truein]{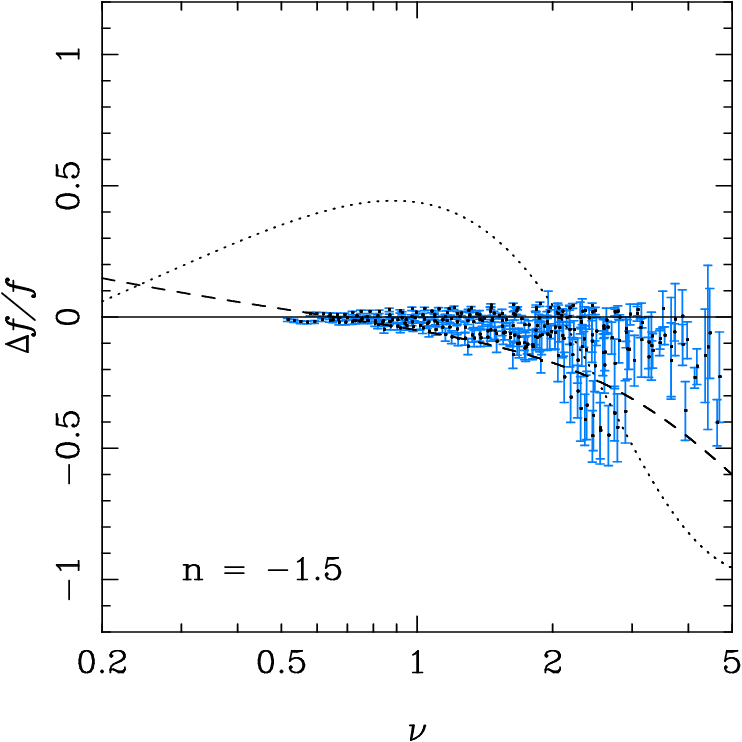} \\
      \includegraphics[width=2.85truein]{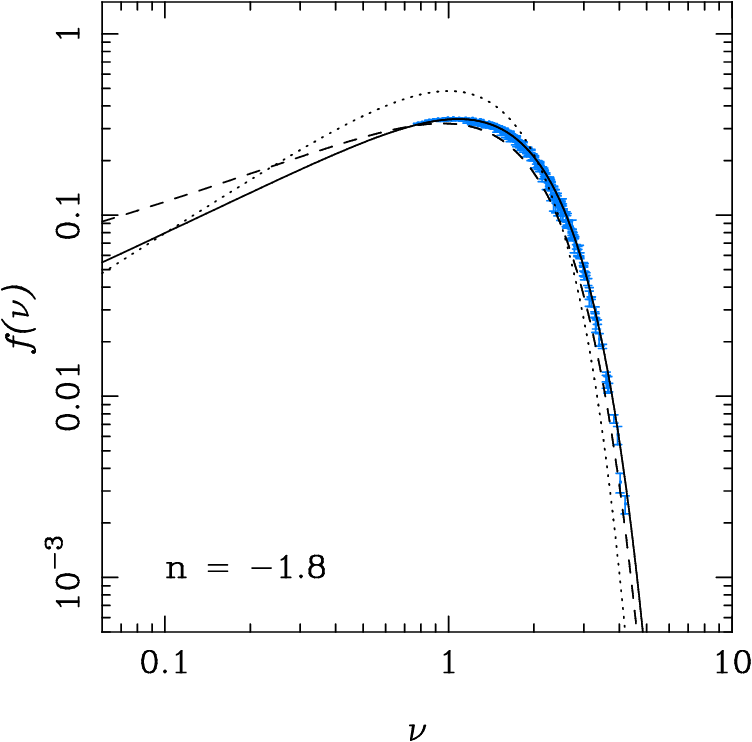} &
      \includegraphics[width=2.85truein]{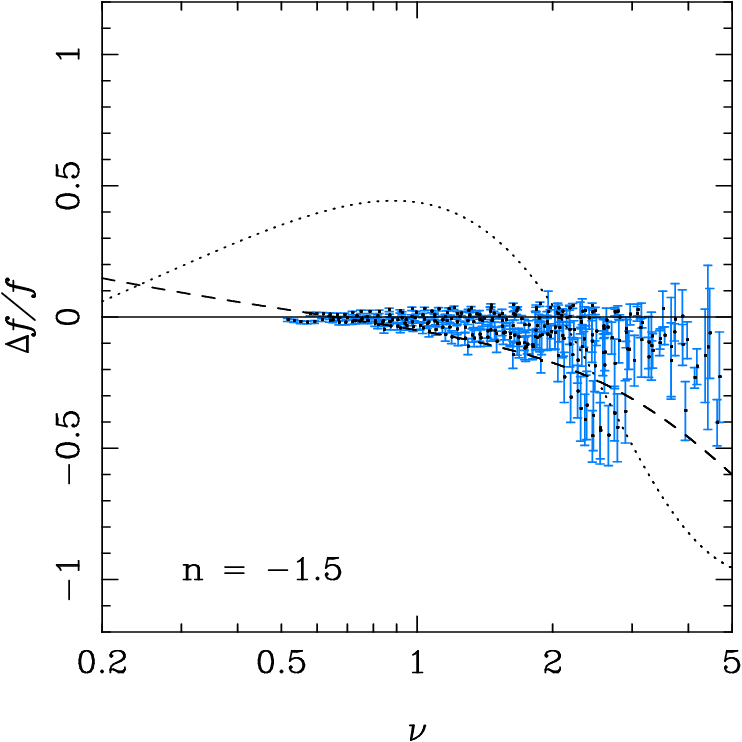} \\
    \end{tabular}
  \end{center}
\caption{Same as Figure (\ref{fig_best_fit1}) for indices $n=1.5$
  (top), and $-1.8$ (bottom).}
\label{fig_best_fit2}
\end{figure*}

Power law models do not have any intrinsic scale apart from the scale of
non-linearity introduced by gravity. 
We can therefore identify an epoch in terms of the scale of
non-linearity $\rnl$.
This is defined as the scale for which the linearly extrapolated 
value of the mass variance at a given epoch $\sigma_L(a,\rnl)$ is
unity.
All simulations are normalized such that $\sigma^2(a=1.0,\rnl=8.0) =
1.0$.  The softening length in grid units is $\epsilon = 0.03 $ in all
runs.

Simulations introduce an inner and an outer scale in the problem and in most
cases we work with simulation results where $L_{box} \gg r_{nl} \geq L_{grid}$,
where $L_{grid}$, the size of a grid cell is the inner scale in the problem.
$L_{box}$ is the size of the simulation and represents the outer scale.

Finite volume effects can lead to significant errors in N-Body
simulations since modes greater than the size of the box are ignored
while generating initial conditions and during evolution
\citep{2005MNRAS.358.1076B, 2006MNRAS.370..993B, 2006MNRAS.370..691P,
  2009MNRAS.395..918B}.
The errors in the mass variance and hence most descriptors of
clustering can become arbitrarily large as the index of the power
spectrum $n$ approaches $-3.0$.
The prescription provided by \citet{2006MNRAS.370..993B} and
\citet{2009MNRAS.395..918B} can be used to find the regime where the
results of a simulation are reliable at a given level of tolerance.
We require that the error in $\sigma^2$ be less than $3\%$ at the
scale of non-linearity.
This requirement severely restricts the level of non-linearity that
can be probed in simulations with indices $n=-2.5$, $-2.2$ and $-2.0$ amongst
the set of models we use here.
We will use these simulations mainly to illustrate the severity of finite box
size effects and compare the mass function obtained in the simulations with
our expectations, but we do not use these simulations for an explicit
determination of the mass function.
In Table (\ref{table_nbody_runs}) we list the power law models
simulated for the present study.
We list the index of the power spectrum $n$ (column 1), size of the
simulation box $N_\rmn{box}$ (column 2), number of particles
$N_\rmn{part}$ (column 3), the scale of non-linearity at the earliest
epoch used in this study (column 4), and, the maximum scale of
non-linearity, $\rnl^\rmn{max}$ (column 6) given our tolerance level
of $3\%$ error in the mass variance at this scale.  
For some models with very negative indices we have run the simulations beyond
this epoch. 
This can be seen in column 5 where we list the actual scale of non-linearity
for the last epoch. 

In the next section we describe 
a procedure to put an upper limit on the high mass bins since in these 
haloes counts are reduced due to finite boxsize considerations, especially at 
late times. 
The counts of haloes in low mass  bins are relatively unaffected 
by finite box considerations. 
We therefore limit errors in the mass function by running the simulation 
up to $\rnl^{max}$ .
Column 7 lists the starting redshift of the simulations for every model. 

Models with a large slope of the power spectrum have more power at small
scales and the relative amplitude of fluctuations at small scales is large. 
Care is required for running simulations of these models as small scales
become non-linear at early times the $r_{nl}$ grows very slowly with the scale
factor. 
A very large number of time steps are
required in order to evolve the system to epochs with a large $\rnl$.
We require that the evolution of the two point correlation function
$\bar\xi$ \citep{1980lssu.book.....P} be strictly self-similar in the 
range of epochs where we use the simulation data. 
This allows us to verify the correctness of evolution.

\begin{figure*}
  \begin{center}
    \begin{tabular}{ccc}
      \includegraphics[width=2.15truein]{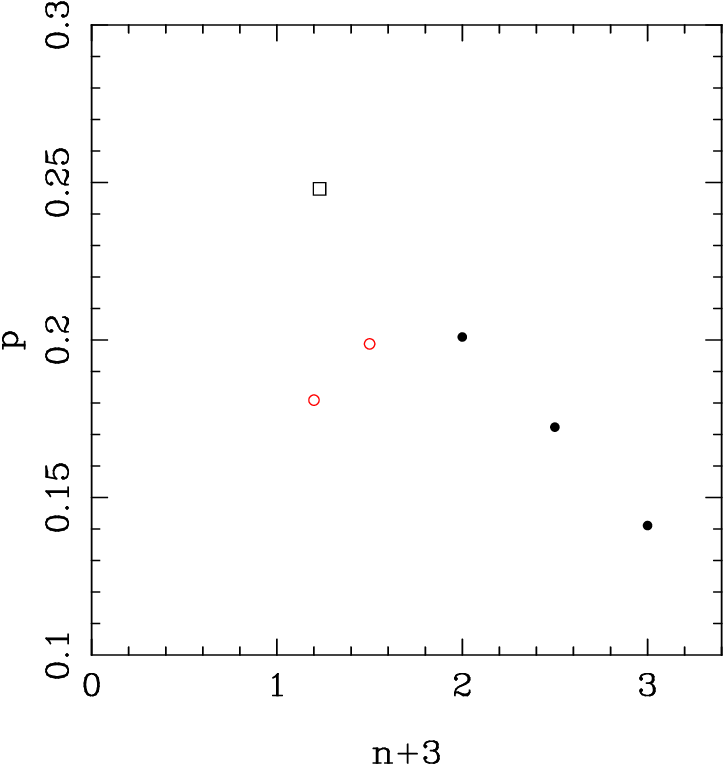} &
      \includegraphics[width=2.15truein]{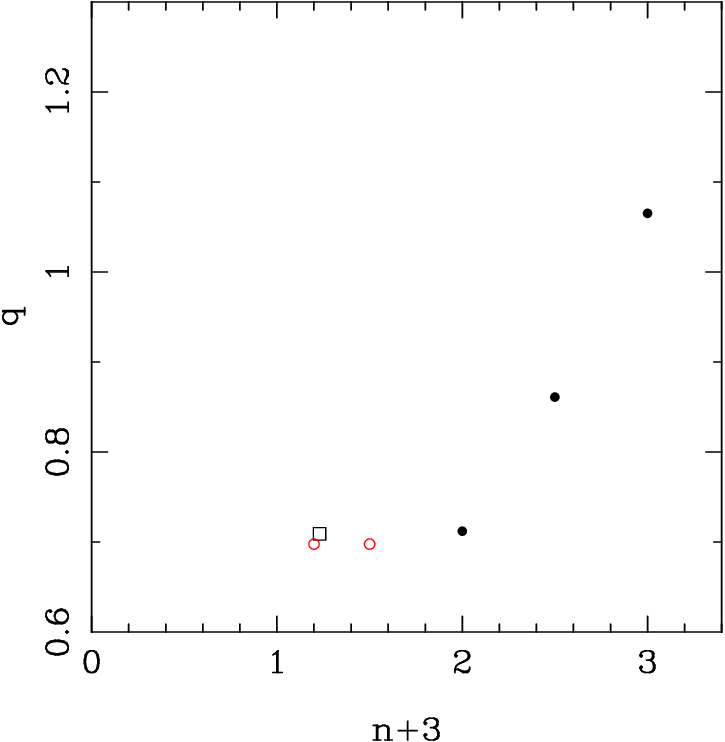} &
      \includegraphics[width=2.15truein]{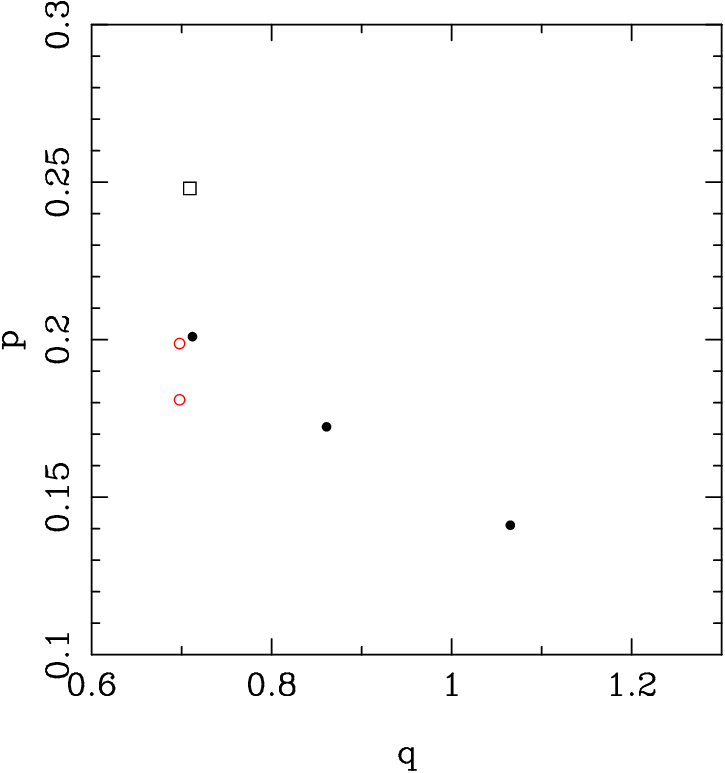} \\
    \end{tabular}
  \end{center}
\caption{Relation between Sheth-Tormen parameters and the index $n$ of
  the power law power spectrum.  Red circles denote data with smaller
  range in $\nu$.  The square denotes the value of
  \citet{2009arXiv0906.1314M}.  See text for details.}
\label{fig_npq_relation}
\end{figure*}

\section {Analysis and Results}
\label{results}

We use the Friends-of-Friends (FOF) \citep{1985ApJ...292..371D}
algorithm with a linking length $l=0.2$ to identify haloes and construct a
halo catalog. 
In order to avoid spurious identification of haloes and also
discreteness noise, we do not use haloes with a small number of
particles --- only haloes with more than $60$ particles are used in our
analysis.

Given the halo catalog one can compute the mass function by first
binning the haloes in mass bins. 
We constructed logarithmic bins in mass with size $\Delta\log m = 0.2$.
Given the halo count per logarithmic mass bin $dn/d\log m$ and 
using the fact that $\bar{\rho}=1$,  we write Equation~\ref{norm} 
\beq
f(\nu) = \frac{6}{n+3}\frac{m}{\bar{\rho}}\frac{dn}{d\ln m} 
\label{eq_nu_powlaw}
\eeq
with $\nu = \delta_c/a\,\sigma(m)$.  
For a power law power spectrum, we have
\beq
\sigma(m) = \left(\frac{m}{m_\rmn{nl}}\right)^{-(n+3)/6}
\label{eq_sigma_powlaw}
\eeq
where $m=4\pi r^3/3$, and we take $\rnl(z=0)\equiv8.0$.  
Note that Equation (\ref{eq_sigma_powlaw}) tells us that it is much
easier to probe the small $\nu$ end of the mass function with larger
indices.
The scale of non-linearity evolves as $\rnl \propto D_+^{2/\left(n+3\right)} =
a^{2/\left(n+3\right)}$, where the second equality follows for the
Einstein-de~Sitter universe.

We choose to fit the Sheth-Tormen mass function of Equation
(\ref{fst}) to our data by the method of $\chi^2$ minimization.
The correspondence with the mass function for ellipsoidal collapse
makes the Sheth-Tormen mass function physically motivated.
The usefulness of this approach is that the best fit values can potentially be
used to compute merger rates, etc.
This however may not be the best choice of the functional form of the mass
function and we comment on it in the following discussion.
The ST mass function has two free parameters, which we denote by $p$ and $q$. 
The condition that all mass must be in haloes provides normalization
for this function (Equation (\ref{norm})). 
This gives
\beq
A=1+\frac{2^{-p}\Gamma(0.5-p)}{\sqrt\pi}
\eeq
as a function of $p$ \citep{2002PhR...372....1C}.

We assume Poisson errors for counts of haloes in a mass bin.
As discussed before, the effect of a finite volume simulation volume
suppresses the count of haloes in the large mass end.
One can either correct it \citep{2007MNRAS.374....2R, 2009MNRAS.394..624R}
or remove these points from the  $\chi^2$  analysis. 
Correcting for these points is a tricky issue since it assumes an
a priori knowledge of the mass function, the quantity which is being 
constructed.  
One can however follow an iterative procedure 
by first starting out with the standard ST mass function or the Press-Schechter 
mass function, then use it to correct the counts at the large mass end of the 
mass function and then do 
the $\chi^2$ analysis to compute  a better ST mass function
and repeat the exercise all over again until one obtains 
a reasonable convergence in the fit.
As we shall see, the dispersion and the goodness of fit does not warrant this
approach and in this paper we choose to remove points affected by more than
$10\%$ in counts (as estimated for the PS mass function) due to box size
effects.   

We begin by fitting the ST mass function to the indices $n=0.0$, $-0.5$,
$-1.0$, $-1.5$ and $-1.8$.  
The raw mass function, i.e., data points, and the best fit ST curve is plotted
in the left column of Figures (\ref{fig_best_fit1}) and
(\ref{fig_best_fit2}). 
The right hand column shows the residuals with respect to the best fit
mass function. 
We also show the PS and the standard ST mass functions in each panel. 
Table (\ref{table_bestfitpq}) shows the best fit values of $p$ and $q$
and the reduced $\chi^2_\rmn{red}$. 

\begin{figure*}
  \begin{center}
    \begin{tabular}{cccc}
      \includegraphics[width=1.6truein]{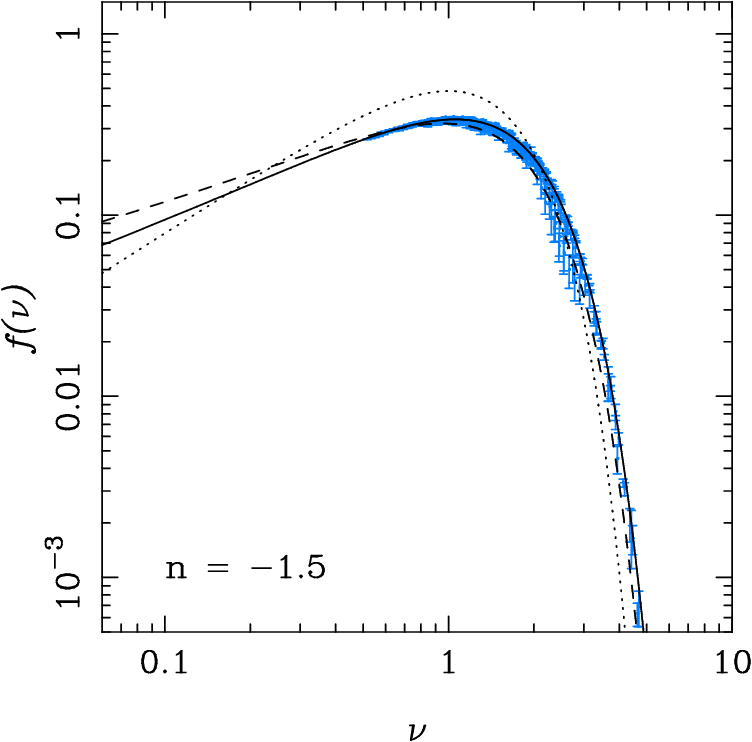} &
      \includegraphics[width=1.6truein]{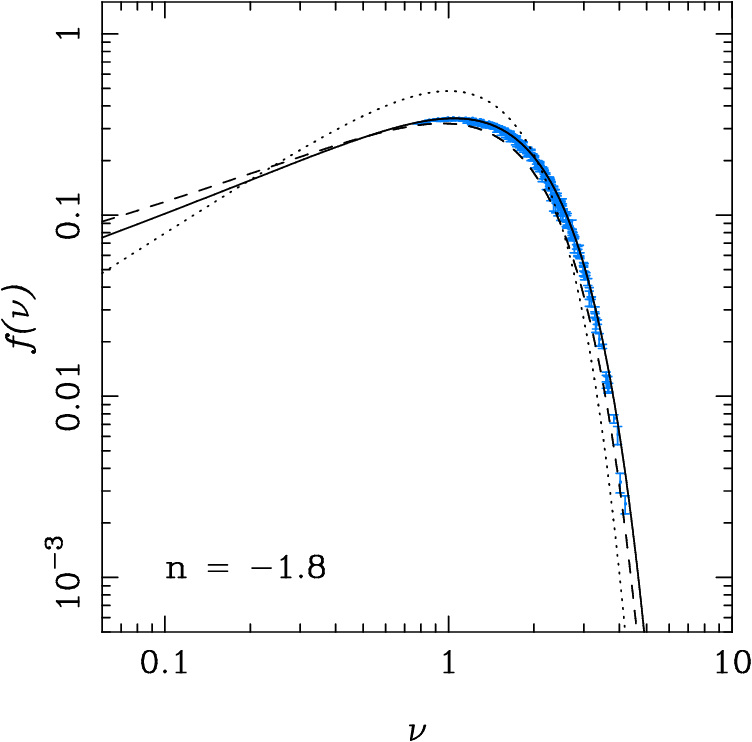} &
      \includegraphics[width=1.6truein]{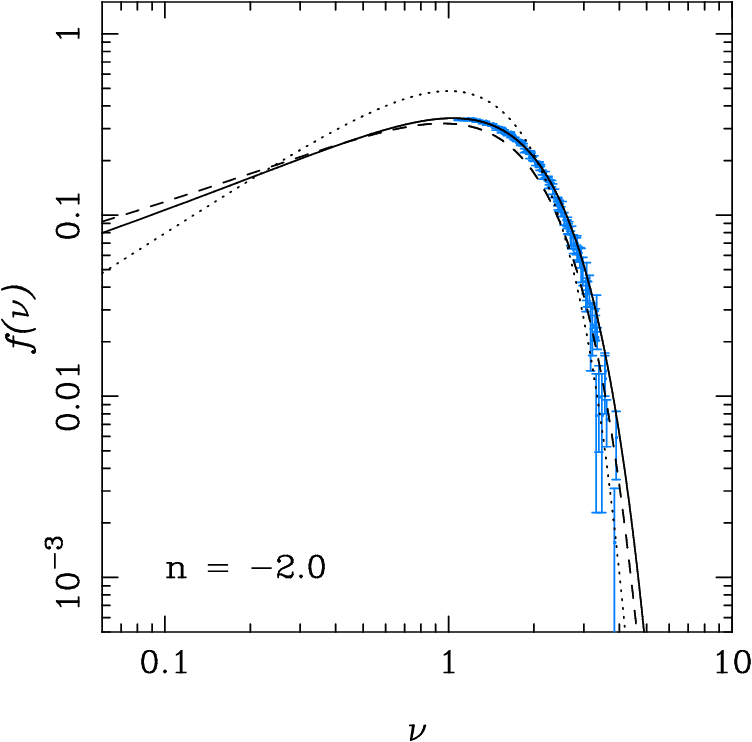} &
      \includegraphics[width=1.6truein]{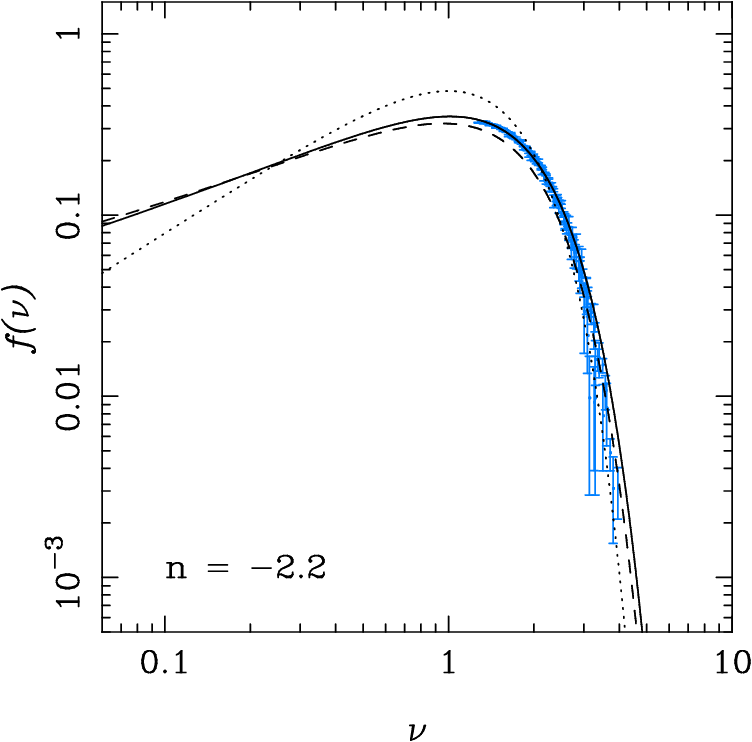} \\
      \includegraphics[width=1.6truein]{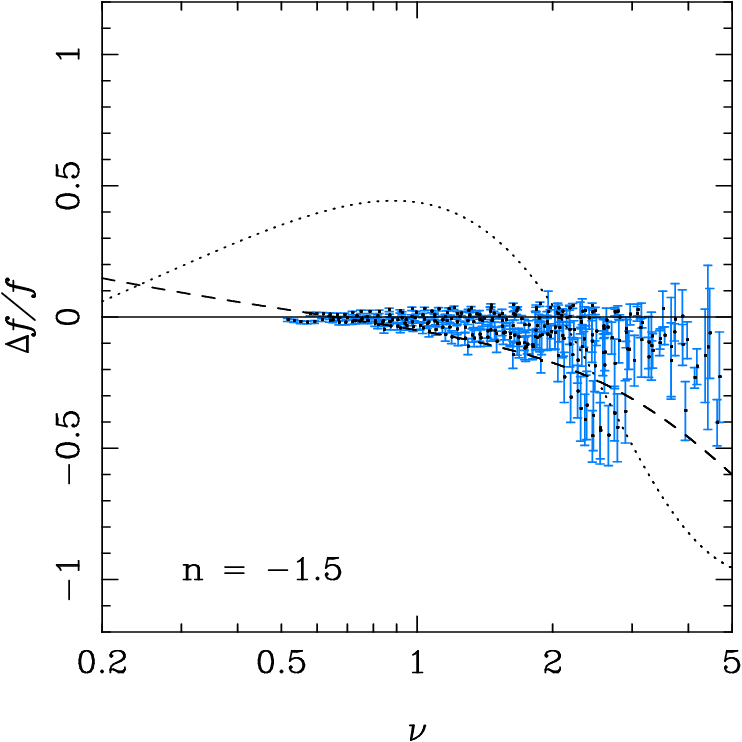} &
      \includegraphics[width=1.6truein]{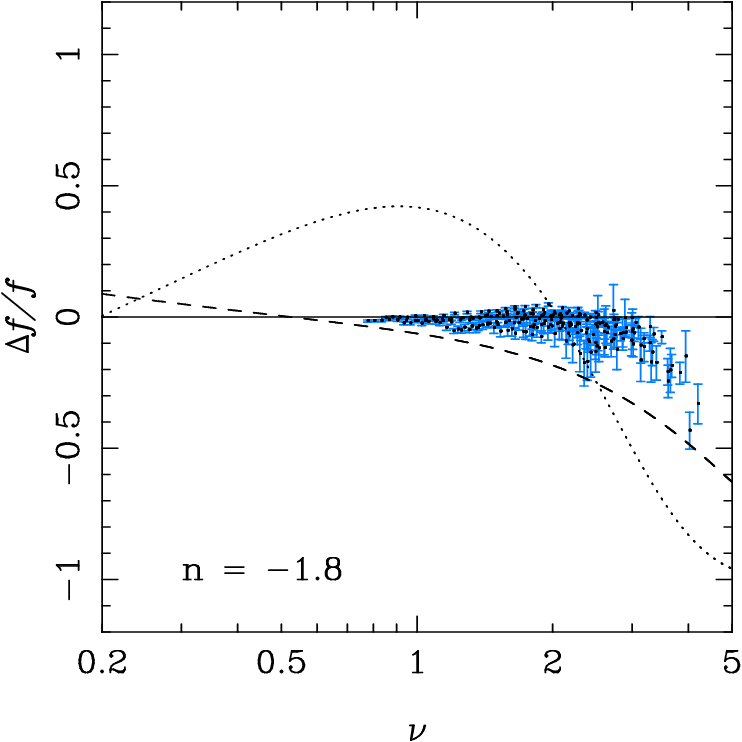} &
      \includegraphics[width=1.6truein]{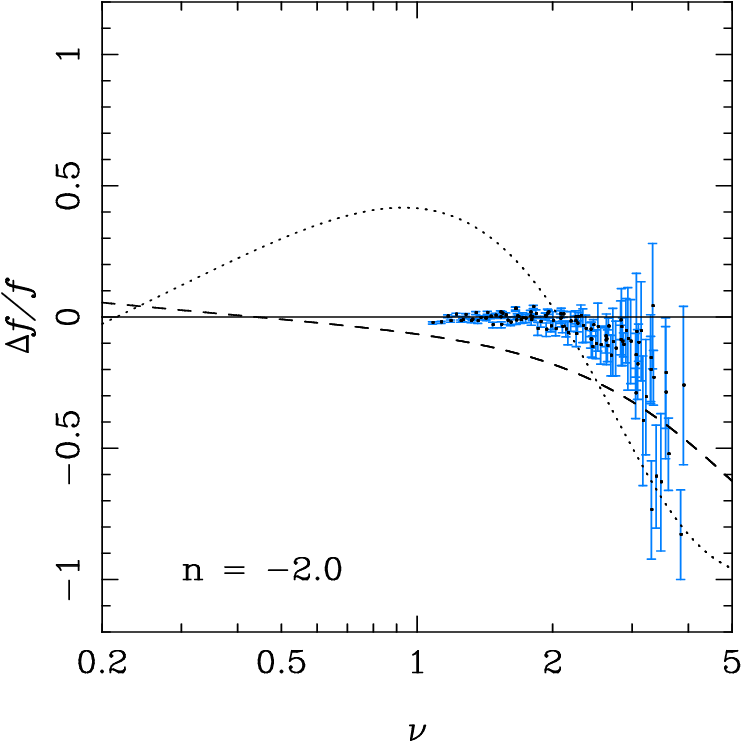} &
      \includegraphics[width=1.6truein]{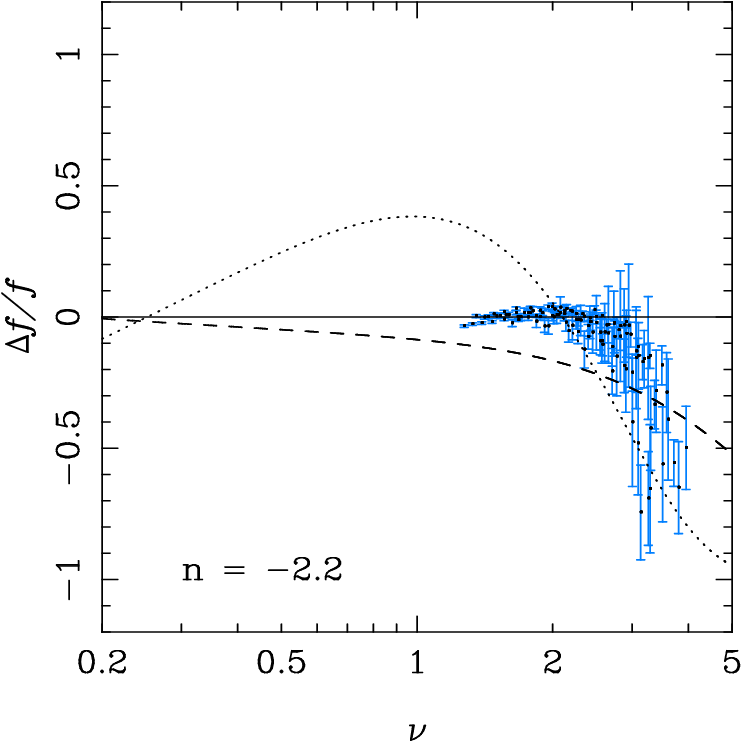} \\
    \end{tabular}
  \end{center}
\caption{Same as Figure (\ref{fig_best_fit2}), except that the solid
  black line here indicates Sheth-Tormen mass function with parameters
  obtained by extrapolating $p(n)$ and $q(n)$ from $-1 \leq n \leq 0$ as shown
  in Figure (\ref{fig_npq_relation}).}  
\label{extrapol}
\end{figure*}

\begin{table}
\caption{Best fit parameters of Sheth-Tormen mass function}
\begin{center}
\begin{tabular}{||r|l|l|l|l|l|l||}
\hline
$n$ & $p$ & $q$ & $\chi^2_\mathrm{red}$ & $p^\mathrm{new}$ & $q^\mathrm{new}$ & $\chi^{2(\mathrm{new})}_\mathrm{red}$ \\
\hline
$0.0$ & $0.141$ & $1.065$ & $4.933$ & $-$ & $-$ & $-$\\
$-0.5$ & $0.172$ & $0.861$ & $3.24$ & $-$ & $-$ & $-$\\
$-1.0$ & $0.201$ & $0.712$ & $31.31$ & $-$ & $-$ & $-$\\
$-1.5$ & $0.199$ & $0.698$ & $4.87$ & $0.232$ & $0.684$ & $5.64$\\
$-1.8$ & $0.181$ & $0.698$ & $4.760$ & $0.250$ & $0.677$ & $6.23$\\
$-2.0$ & $-$ & $-$ & $-$ & $0.262$ & $0.677$ & $3.94$\\
$-2.2$ & $-$ & $-$ & $-$ & $0.274$ & $0.698$ & $4.56$\\
\hline 
\label{table_bestfitpq}
\end{tabular}
\end{center}
\end{table}

\section{Discussion}
\label{discussion}

At the outset a visual inspection of Figure (\ref{fig_best_fit1})
shows a clear trend in the shape of the mass function with index $n$
of the power spectrum.
We also see that the evolution is self-similar across different epochs
for all indices. 
This is as expected for power law models in Einstein-de~Sitter
background.

In Figures (\ref{fig_best_fit1}) and (\ref{fig_best_fit2}) we find
that for the more negative indices $n=-1.8$, $-1.5$, $-1.0$ we have a 
smaller dynamic range in $\nu$ as compared to $n=-0.5$, $0.0$.
This is expected given that we have not evolved simulations with these
indices over a large range of epochs due to finite boxsize considerations.
Since the slope of $f(\nu)$ at the small $\nu$ end is related to
$p$ --- it is $f(\nu) \propto \nu^{1-2p}$ as $\nu\rightarrow 0$ --- one
cannot trust these results as much as the results with $n=-0.5$ and
$0.0$ which probe $f(\nu)$ out to much smaller values in $\nu$. 
The best fit values of $(p,~q)$ and $\chi^2_{\rmn{red}}$
are shown in columns 2, 3 and 4 of Table~\ref{table_bestfitpq} for 
every model (column 1).
Surprisingly $\chi^2_{\rmn{red}}$ is relatively low for all indices as
compared to the index $n=-1.0$. 
Part of this is due to several outliers for the $n=-1.0$ spectrum which have
more than $10\sigma$ deviations. 
Removing these points we find that $\chi^2_{\rmn{red}} = 11.3$ from 
the original value of $\chi^2_{\rmn{red}} = 31.6$. 
The corresponding values of $(p,~q) $ changes to $(p,~q) = (0.22,~0.73)$ 
from  $(p,~q) = (0.20,~0.71)$.
In Figure (\ref{fig_npq_relation}) we also add the best fit value of
$p$ and $q$ from the recent work of \citet{2009arXiv0906.1314M}, which
used $49$ realizations of an LCDM simulation with $640^3$ particles in a
box of side $L_\mathrm{box} = 1280 h^{-1}$Mpc.
We identify the effective index at $z=0$, by computing 
\beq
\left.\frac{d\log\sigma(m)}{d\log m}\right|_{\sigma=1} = 
-\frac{(n_\rmn{eff}+3)}{6}
\eeq 

Next we look at the dependence of the two ST parameters $(p,~q)$ on the index
of the spectrum $n$ and also their interdependence.  
In Figure (\ref{fig_npq_relation}) we find that all the trends seen
between $p$, $q$ and $n$ ($n_\rmn{eff}$ for LCDM) are smooth when one considers
the LCDM model and the power law models with indices $n=0.0, -0.5,
-1.0$.
We find that the relation between $p$ and $n$ is approximately 
\beq
p(n) \simeq -0.0605~n + 0.141
\label{eq_pn}
\eeq
The trends in these parameters show deviations when one adds the models
$n=-1.5$ and $-1.8$, especially in the values of $p$; deviations in
$q$ are not as drastic.
However if we use Equation (\ref{eq_pn}) and fix $p$ for the indices
$n=-1.5, -1.8$ and redo the $\chi^2$ analysis to fit $q$ then the
$\chi^2$ does not change much, thereby corroborating our argument that
the values of $p$ for the models $n=-1.5$, $-1.8$ are unreliable due to the
small range in $\nu$. 
The new values of $p^\mathrm{new}, q^\mathrm{new},
\chi^{2(\mathrm{new})}_{\mathrm{red}}$ are listed in columns 5-7 of Table
(~\ref{table_bestfitpq}).  
Figure~(\ref{extrapol}) plots the best fit 
ST curve using the extrapolated value of $p$ for the models $n=-1.5, -1.8$,
$n=-2.0$ and $-2.2$. 
Since we have not probed highly non-linear scales for these models 
the range of $\nu$ is even more limited, so much so that 
a full $\chi^2$ minimization for $(p,~q)$ is futile.
We therefore do not list the best fit values of  $(p,~q)$
in Table (~\ref{table_bestfitpq}) and proceed to 
do a similar fitting  using Equation  (\ref{eq_pn}) 
for the models $n=-2.0$, $-2.2$ and find that 
$\chi^{2(\mathrm{new})}_{\mathrm{red}}$ is is in the same range as for other
models. 
The values 
 $p^\mathrm{new}, ~q^\mathrm{new}, ~\chi^{2(\mathrm{new})}_{\mathrm{red}}$
for these models are given in columns 5-7 of Table
(~\ref{table_bestfitpq}). 
Again we find that the $\chi^{2(\mathrm{new})}_{\mathrm{red}}$
is reasonable giving credence to the extrapolation of Equation  (\ref{eq_pn}).

However the trend between $p,~q$ and $n_\rmn{eff}$ is reversed when we
compare with the LCDM runs of \citet{2009arXiv0906.1314M}.
Here $p$ decreases and $q$ increases with increasing redshift
(increasing $n_\rmn{eff}$), at $z=0.5$. 
It is important to note that in the simulations used in
\citet{2009arXiv0906.1314M}, it is not possible to probe $\nu < 1$. 
However at this redshift, $\Omega_{\Lambda}$ has a dominant role
thereby making a simple interpretation difficult.
We are carrying out a series of numerical experiments where we simulate power
law models in a background cosmology with a cosmological constant to develop
further understanding of this issue.

\citet{2006ApJ...646..881W} reported a reduced $\chi^2_\rmn{red}\sim
5$ for the Sheth-Tormen mass function with their data.
Our values of $\chi^2_\rmn{red}$ are similar and confirm that the
Sheth-Tormen mass function is inadequate in fitting the data in power
law models. 

The Sheth-Tormen parameters also show a clear dependence on the index
of the spectrum.
One approach of alleviating this problem would be to better model
barrier shape using simple models like the ones we have here.
These can then be used to construct and compare with more complicated
barriers arising due to different cosmology and a scale dependant
index like the CDM class of models.
The other approach can be purely phenomenological: one can use results
from simpler models like those studied here and understand how further
complications in the model, like running index and different
cosmologies, affect the shape of the mass function.

Our focus has been on demonstrating that the mass function has a
dependence on the slope of the power spectrum.
In this study we have not looked at how a change in the definition of
the halo---for example, change in the linking length in FOF halo
finder or using the SO halo finder instead of FOF
\citep{1994MNRAS.271..676L, 2008ApJ...688..709T}---affects the mass
function.
Indeed one expects that the halo definition changes the amplitude
of the mass function \citep{2008MNRAS.385.2025C, 2002ApJS..143..241W,
  2008ApJ...688..709T} but we do not expect it to affect our results in a
significant manner.
We postpone a study of the dependence of the Sheth-Tormen parameters
on halo definition to a later paper.

At least one mass function model with dependence on spectrum index
have been considered in the literature \citep{2001NYASA.927....1S}.
This is derived by modifying the peaks picture of
\citet{1986ApJ...304...15B}, in which the initial Gaussian density
field is filtered at a certain scale and number
density of peaks of a certain height in the smoothed field is taken to
be the number density of haloes of mass corresponding to the smoothing
scale.
\citet{2001NYASA.927....1S} modifies this mass function by smoothing
the density field with a range of filter sizes and identifying haloes
with peaks of varying heights so that more massive haloes
correspond to higher peaks.  
For a power law spectrum of initial density fluctuations, this mass
function can be derived analytically and has an explicit dependence on the
spectral index. 

We compare the \citet{2001NYASA.927....1S} mass function with our
simulations by using a Gaussian filter instead of real-space top hat
in Equation (\ref{norm}).  
We find that the model does not fit our results well.  

\section{Conclusions}
\label{conclusions}

We summarize the conclusions of the present study here.
\begin{itemize}
\item 
We find that the mass function is not universal and has an explicit dependence
on the power spectrum.
\item 
The Sheth-Tormen parameters show a systematic trend with index of the power
law power spectrum.  
We also find that there is a correlation between these parameters.
\item 
Evolution of the mass function $f(\nu)$ in power law models in an Einstein
de~Sitter background is self-similar, i.e., the functional form does not
change with time for a given power law model. 
We do not expect this in the CDM class of spectra since the effective index,
$\neff$, changes with redshift. 
However, the spectrum dependence of $p$ and $q$ is not very strong, and the
range over which $\neff$ varies is not very large, hence the usual ST
parameters are an adequate first approximation.
Variation of ST parameters becomes relevant if we require a very precise
description of the mass function.
\item 
Our $\chi^2$ analysis shows that Sheth-Tormen mass function is inadequate and
that better modeling of collapse with more parameters is needed. 
\item 
Finite box size considerations impose serious limitations on
running simulations with large negative indices of the power
spectrum.  This end of the spectrum probes the mass function of high
redshift objects.  The way out would be to run orders of magnitude
larger simulations for indices with $n\rightarrow -3$.  This is a
challenge and would be overcome only in the next generation of
simulations.
One can possibly use zoom in simulations but the displacements for these
models are large, as is the large scale tidal field, and hence the extent to
which one can benefit from zoom in simulations is not obvious.
\item
For models with $n \gg -3$, the rate of growth for $\rnl$ with time is very
slow and as a result a large number of time steps are required for evolving
the system. 
The Adaptive TreePM \citep{2009MNRAS.396.2211B} may be useful for running such
simulations.  
\end{itemize}

\section*{Acknowledgments}

Computational work for this study was carried out at the cluster
computing facility in the Harish-Chandra Research Institute
(http://cluster.hri.res.in/index.html).
This research has made use of NASA's Astrophysics Data System.
The authors would like to thank Ravi Sheth for useful discussions.

\label{lastpage}
\end{document}